\RequirePackage[2020-02-02]{latexrelease}
\documentclass[prl,showpacs,preprintnumbers,amsmath,amssymb]{revtex4}

\usepackage[dvips]{graphics}
\usepackage{tikz}

\begin{document}

\title{Pionic Entanglement in Femtoscopy: A Lesson in Interference and  Indistinguishability}

\author{Vlatko Vedral}
\affiliation{Clarendon Laboratory, University of Oxford, Parks Road, Oxford OX1 3PU, United Kingdom}

\date{\today}

\begin{abstract}
\noindent We present an analysis of recent experiments in femtoscopy by the STAR collaboration in terms of the protocol of entanglement witnessing involving purity measurements. 
The entanglement is between the charge and momentum degrees of freedom of pions and the state purity measurements ultimately rely on the bosonic nature of the detected pions. 
The pion experiment is intended to measure the size of nuclei and the distance between the nuclei involved, however it indirectly confirms that the states of differently charged pions are entangled through an entanglement witness based on the purity of various pionic states. The entangled state of pions can be modelled straightforwardly dynamically using a simple Hamiltonian. Quantum indistinguishability plays a key role in this analysis and we make comparison with the equivalent photonic experiments. 
\end{abstract}


\maketitle                           

The Hanbury-Brown Twiss (HBT) intensity interferometer \cite{HBT} was invented in order to measure the angular sizes of various astronomical objects like stars. As the name suggests, the key idea is to look at the intensity-intensity correlations between two different detectors receiving light from the same astronomical object. The advantage of the intensity (as opposed to the amplitude) correlations is that they are more robust to fluctuations and therfore more readily accessible. But why would intensities interfere? We normally think of amplitudes interfering, and not the probabilities, which arise by mod squaring the amplitudes. However, interference can take place between intensities for the simple reason that superpositions could also contribute to them. This is clear from the classical wave theory, but it is even easier to understand quantum mecanically. Here we will mainly follow the quantum explanation (not because it is easier, but) simply because it is more fundamental, leading - as it does - to the classical one in a special limit of large amplitude coherent states.
 
Quantum mechanically, the intensities interfere because of the fact that identical particles, when they are in indistinsguihable states, tend to bunch (bosons) or anti-bunch (fermions). Let us focus on bosons since this is the kind of interference we will be describing in this paper (photons, which are relevant for stellar interferometers and pions which are relevant for nuclear femtoscopy are both bosonic particles).

The reader may find it surprising that the HBT idea can also be used in high energy physics to probe nuclear distances, see e.g. \cite{Baym,Lisa}. The principle is the same as in the stellar interferometer and we beriefly review its role in femtoscopy before proceeding to show how it also acts as an entanglement witness. We can think of various high energy nuclear processes as resulting in the emission of pions. In the crudest of approximations, we could represent states of the pions as plane waves. These plane waves are emerging from different parts of the nucleus would then interfere. If we are talking about the pions that have the same charge, then there is an amplitude for two pions to arrive at two detectors in two different ways. 
\begin{equation}
A \propto e^{ik_1x_1} e^{ik_2x_2} + e^{ik_1x_2} e^{ik_2x_1}
\end{equation}
(there should be four different paths in general, but we are assuming that they are pairwise equal; no generality is lost). 
Given that they are indistinshuishable, we must add up the amplitudes. The interference between these two is the HBT interference and it is an intensity interference as it is to do with the number states of pions. Clearly, the amplitude above is exactly what one would get by doing a completely classical treatment (and this is not sursrising as quantum physics does reproduce the classical wave behaviour in a certain, well-defined, limit). The full quantum calculation would simply have to evolve the pion creation and annihilation operators and then compute the expected values of the resulting number operators (at $x_1$ and $x_2$).  We will show how this is done using a different variant of the HBT experiment. 

When we calculate the intensity $I=|A|^2$, the relevant terms are the cross terms $\cos \delta k \delta x$. If we have an extended source, then we have to sum up over all the relevant contributions, and the coherent contribution becomes the Fourier transform of the source (the van-Citter Zernike theorem \cite{Born}). In the simplest instance the source could be represented by a top hat function and we then obtain a simple form of the coherence:
\begin{equation}
C = \frac{\sin^2 (k\alpha b/2)}{(k\alpha b/2)^2}
\end{equation}
where $k=2\pi/\lambda$, $b$ is the distance between the detectors and $\alpha$ is the angular size of the nucleus (or the star in the case of the original HBT).
Therefore, by measuring $C$ (and knowing $\lambda$ and $b$) we can then estimate the angular size of the nucleus and therefore its diameter (since we know the distance to the detectors). 

The recent experiment reported in \cite{STAR} is a variant of HBT. It involves emission of entangled pairs of pions which are then detected by two detectors. Each detector detects two pions,
and the possibilities are that detector A detects two positive pions (B therefore detects two negative ones), vice versa, and finally both detectors detect one positive and one negative pion. It is this last possibility that has two indistinguishable ways of happening, which is where the interference occurs. The amplitude for this is given by
\begin{equation}
A \propto e^{ik_1x_1} e^{ik_2x_2}e^{ik_3x_1} e^{ik_4x_2}+ e^{ik_1x_1} e^{ik_4x_1}e^{ik_2x_1} e^{ik_3x_2}
\label{amp}
\end{equation}
Due to the momentum conservation, there are restrictions on ks, which we will discuss below. Suffice it to say here that a full treatment will lead to the coherence that behaves as
\begin{equation}
C = \frac{\sin^2 (k\alpha b)}{(k\alpha b)^2}\cos^2 (k\beta b)
\end{equation}
where $\alpha$ is the angular distance between the two sources, and $\beta$ is the angular width of each of them. This is again a Fourier transform of the convolution of the top hat function (representing each nuclei) with two delta functions (represeting the locations of the nuclei). Note that it contains both the informaiton about the angular size of nuclei as well as the angular distance between them. 

In order to show how this experiment is, in fact, a witness of the entanglement between pions, we will now present the full quantum analysis. The brief details of the experiment are as follows. Two gold nuclei are scattered at high energies off one anther. This results in each emitting a rho particles, each of which subsequently decays into a positive and a negative pion. It is these pions that are ultimately detected and produce interference as outlined above.

In order to write down the state of two pions, we need a quantum model for this process \cite{Gyulassy}. The process of the annihilation of the rho particle and the creation of the two pions is represented by the following Hamiltonian
\begin{equation}
H = -g \int dx^3 \psi^{\dagger} (x,t)\psi (x,t) \phi (x,t)
\end{equation}
where $g$ is the strength of the interaction, $\phi (x,t)=\phi^+ (x,t)+\phi^- (x,t)$, $\psi (x,t)=\psi^+ (x,t)+\psi^- (x,t)$ and 
\begin{eqnarray}
\phi (x,t)^+ = \sum_k c_k e^{ik_\mu x^\mu} \nonumber\\
\phi (x,t)^- = (\phi (x,t)^+)^{\dagger}\nonumber \\
\psi (x,t)^+ = \sum_k a_k e^{ik_\mu x^\mu} \nonumber\\
\psi (x,t)^- =\sum_k b^\dagger_k e^{ik_\mu x^\mu}\nonumber\\
(\psi (x,t)^\dagger)^+ =  (\psi (x,t)^-)^{\dagger} \nonumber\\
(\psi (x,t)^\dagger)^- = (\psi (x,t)^+)^{\dagger}
\end{eqnarray}
where $k_\mu x^\mu = kx-\omega t$. Here the $c$ operators annihilate the rho meson, the $a$ operators annihilate the positive pion and $b$ the negative pion (and the hermitian conjugates create them). 
This Hamiltonian is the high energy analogue of the down-conversion process in quantum optics, where a single photon is converted into two photons through a non-linear medium. 
The Hamiltonian above represents $8$ different processes (in which rho can be created or destryed and pions can also be created and destroyed), but only the term, $ca^{\dagger}b^{\dagger}$, is relevant. When we start from the state containing one rho particle and apply this Hamiltonian, we obtain a superposition of the initial state and the state with no rho particles and containing an entangled state of the pions 
\begin{equation}
|\Psi\rangle = c_0 |\rho, 0,0\rangle + c_1 \int d\omega_1d\omega_2dq_1dq_2 f(q_1,q_2,\omega_1,\omega_2)|0,q_1\omega_1,q_2\omega_2\rangle 
\end{equation} 
subject to the condition that $q_1+q_2 = q_\rho$ and $\omega_1 + \omega_2 = \omega_\rho$ (which itself is part of the definition of the function $f$, whose exact form is not of interest to us here). The exact form of the amplitudes is also not directly relevant as only the term containing pions will contribute to interference. 

The interference term, where each detector detects one positive and one negative pion, is then given by the following $8$-point correlation:
\begin{equation}
\langle \psi^- (y,t)(\psi^\dagger)^- (y,t)\psi^- (x,t)(\psi^\dagger)^- (x,t)(\psi^\dagger)^+ (y,t)\psi^+ (y,t) (\psi^\dagger)^+ (x,t)\psi^+ (x,t)\rangle
\end{equation}
where the average is taken with respect to the second term in $|\Psi\rangle$ since, as we noted, this represents the only contribution to the detection of pions (the rest constitutes the detection of the rho meson). 
This expression would in quantum optics be known as the $g^{(4)}$ coherence \cite{Glauber} (where instead of the $\psi$ operators we would have the 
positive and negative frequency electric field operators). This quantity is basically the probability to observe a positive and a negative pion in each detector.
The reason why this will lead to interference is simple to see when the operators are expanded in the momentum basis
\begin{equation}
\langle a_p b_q a_r b_s a^\dagger_n b^\dagger_m a^\dagger_k b^\dagger_l  \rangle  = \langle a_p b_q a_{q_\rho-q} b_{q_\rho-p} a^\dagger_p b^\dagger_q a^\dagger_{q_\rho-q} b^\dagger_{q_\rho-p}  \rangle + \langle a_p b_q a_{q_\rho-q} b_{q_\rho-p} a^\dagger_{q_\rho-q} b^\dagger_{q_\rho-p} a^\dagger_p b^\dagger_q  \rangle
\end{equation}
where we have use the fact that the momentum states are correlated because of momentum conservation and the fact that all other terms must vanish since different number states are orthogonal to each other. Because of entanglement \cite{Vedral}, namely the fact that pions populate modes whose momenta are correlated, the $g^4$ coherence behaves effectively the same as $g^2$ for bosonic number states. 

The expression will reproduce the result in eq.(\ref{amp}). Note that we are neglecting the electromagnetic interactions
between the pions. A complete treatment should, of course, include the attraction between the like pions and the repulsion otherwise, but we are here interested in the dominant effect only, which is due to the particle statistics. 

The other two detection options, which is that two pions of positive charge are detected in one detector and two of negative in the other, each can be assigned a similar
expression (though they only add up incoherently). In the fully entangled state of pions the interfering term has a probability $1/2$ and each incoherent term has
a probability of $1/4$. Here therefore the reduction in coherence is a direct consequence of the fact that in half of all cases the terms are fully distinguishable - since they contain different charges - and do not lead to interference. Given this, let us now explain why this procedure constitutes an entanglement witness. 

The analysis will be much simpler if we use the qubit notation, so that a positive pion corresponds to $|0\rangle$ and a negative pion to $|1\rangle$ (we are then ignoring the spatial degrees of freedom as well as the bosonic nature of the pions since qubits are fully distinguishable). The state of two 
entangled pions is then $|\Psi^+\rangle = |01\rangle + |10\rangle$ (not normalized). The above experiment is based on detection of two such states $12$ and $34$ at a time. The relevant probabilities for interference are given by 
\begin{equation}
p = tr \{ P_{13}\otimes P_{24} (\rho_{12}\otimes \rho_{34})\}
\end{equation}
where $P$ is the projection onto the symmetric state of the respective pions (labelled by the subscripts). It is clear that if the state of pions was not entangled (but, say, just a product of the states $\pi^+$ and $\pi^-$, $|01\rangle\otimes |01\rangle$), the HTB interference would not occur. In fact, the entanglement witness here is similar to the one based on puruty (or linear entropy) \cite{Horodecki}. By measuring projections onto the symmetric and antisymmetric subspaces we can calculate the total purity of the state as well as the local purities of the subsystems \cite{Bovino}. If the total purity exceeds both of the local ones, the total state must then be an entangled one. This is intuitively clear since for maximally entangled states the total purity is one, while the local ones are both zero (since the reduced states are equal mixtures of $|0\rangle$ and $|1\rangle$). 

The qubit version of the high energy experiment is best understood by expanding the two entangled states (entangled state $12$ and another one $34$) in the basis of the detected qubits  $13$ and $24$:
\begin{equation}
|\Psi_{12}^+\rangle |\Psi_{34}^+\rangle = |00\rangle_{13}  |11\rangle_{13} +  |11\rangle_{13} |00\rangle_{13} + |\Psi_{13}^+\rangle |\Psi_{24}^+\rangle - |\Psi_{13}^-\rangle |\Psi_{24}^-\rangle
\end{equation}
where $|\Psi^-\rangle = |01\rangle - |10\rangle$ (which is a state that is antisymmetric and does not occur with pions which are bosons). 
It is now clear that the interference comes from the last two terms where different entangled states are themselves entangled. Entangled entanglement  \cite{Zeilinger} is therefore behind the higher order coherences in quantum mechanics and femtoscopy analysed here is just one of many places where it plays a crucial role.  

The interplay between spatial and internal degrees of freedom based on particle statistics is, of course, well known in quantum information in general. Protocols such as entanglement swapping \cite{Omar} and entanglement distillation \cite{Paunkovic} can be performed by only exploiting the boconic (fermionic) nature of the qubits used.  In this vain, the above pionic high energy experiment can be viewed as the confirmation of the spatial entanglement of pions based on the detection of their charge properties (which could be thought of as an internal degree of freedom). One wonders if there are further insights to be gained by cross-pollination between quantum information and high energy physics.

\textit{Acknowledgments}: VV is grateful to the Moore Foundation and the Templeton Foundation for supporting his research.

\end{document}